\documentclass{v19cosmo_MY}

\usepackage{amsmath}
\usepackage{amssymb}
\usepackage{cite}

\def\SU{\mathrm{SU}}

\newcommand{\Photo}{\includegraphics[height=30mm]{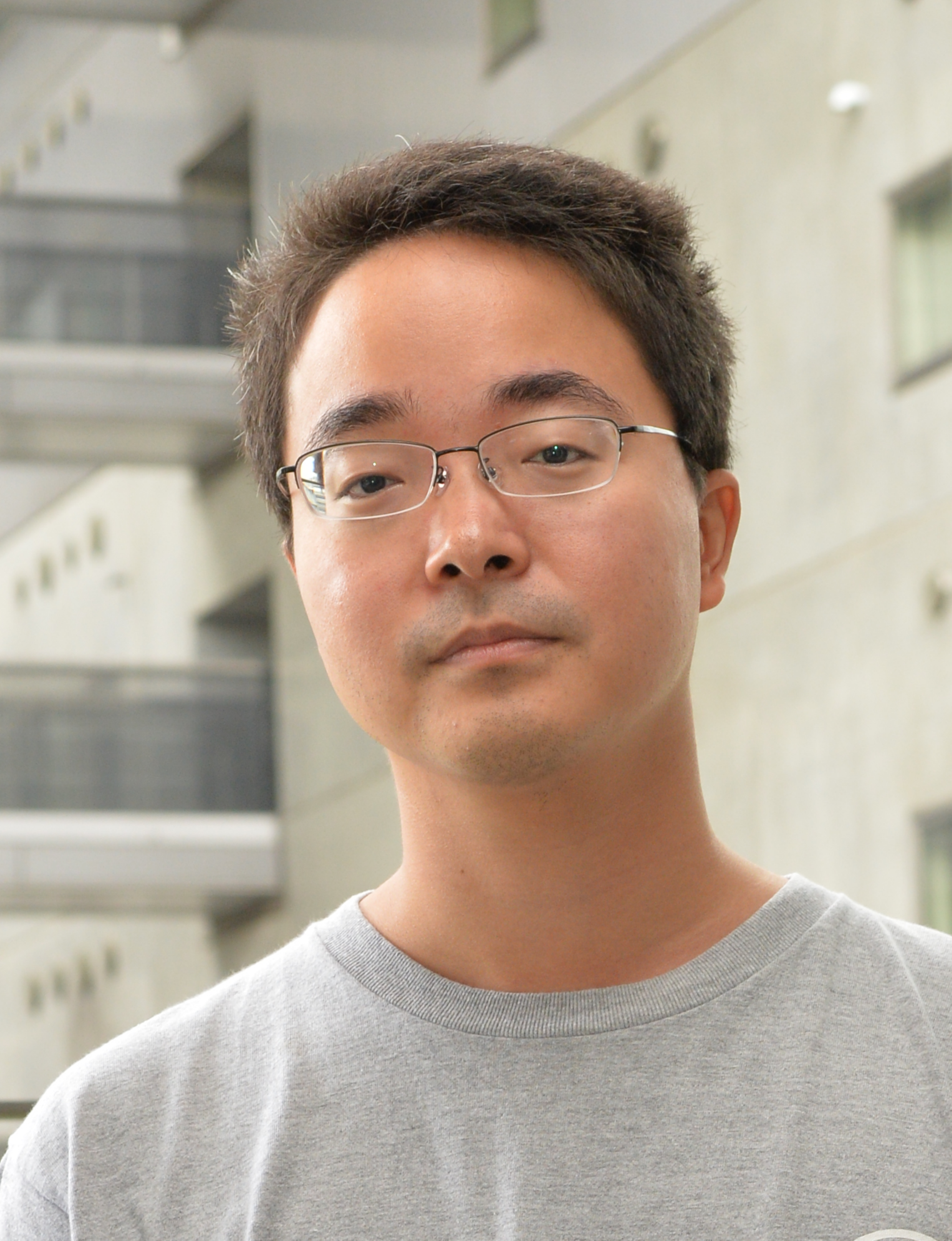}}

\begin{document}
\vspace*{4cm}
\title{Electroweak Quintessence Axion as Dark Energy\footnote{Contribution to proceedings for Rencontres du Vietnam, August 2019.}}

\author{Masahito Yamazaki}

\address{Kavli Institute for the Physics and Mathematics of the Universe (WPI),  UTIAS, \\
The University of Tokyo, Kashiwa, Chiba 277-8583, Japan}

\maketitle\abstracts{We discuss the electroweak quintessence axion as a candidate for dark energy, taking into account
observational as well as quantum-gravity constraints.}

%%%%%%%%%%%%%%%%%%%%%%%%%%%%%%%%%%%%%%%%%%%%%%%%%%%%%%%%%%
\section{Dark Energy}
%%%%%%%%%%%%%%%%%%%%%%%%%%%%%%%%%%%%%%%%%%%%%%%%%%%%%%%%%%

It has now been established that we live in the Universe where
the dark energy accounts for a significant fraction of the 
energy density of the Universe.  
One of the puzzles for dark energy is that 
the energy scale of the dark energy, which we denote by $\Lambda_0$,
is much smaller than the cutoff scale of the theory:\begin{align}
\Lambda_0^4\simeq 10^{-120} M_{\rm Pl}^4 \simeq O(\textrm{meV})^4 \ll M_{\rm Pl}^4 \;.
\label{Lambda_0}
\end{align}
Here we have chosen the cutoff scale to be 
the reduced Planck scale $M_{\rm Pl}\simeq 2\times 10^{18} \textrm{GeV}$, where quantum gravity effects are expected to be significant.

The goal of this short note is to summarize a scenario where such a small scale is naturally realized,
based on the author's recent paper \cite{Ibe:2018ffn} with M.~Ibe and T.T.~Yanagida.

Since \eqref{Lambda_0} is formulated as a tension between IR physics (dark energy scale $\Lambda_0$)
and UV physics (cutoff scale $M_{\rm Pl}$), any understanding of the hierarchy \eqref{Lambda_0}
will inevitably involve physics at the cutoff scale.
In this spirit, in the following we will take into account swampland constraints \cite{Vafa:2005ui,Ooguri:2006in} (see e.g.\ \cite{Yamazaki:2019ahj} for a short note) originating from quantum gravity.

%%%%%%%%%%%%%%%%%%%%%%%%%%%%%%%%%%%%%%%%%%%%%%%%%%%%%%%%%%
\section{Quintessence}
%%%%%%%%%%%%%%%%%%%%%%%%%%%%%%%%%%%%%%%%%%%%%%%%%%%%%%%%%%

One common scenario for dark energy is that it is a cosmological ``constant''.
Since there are no free adjustable parameters in quantum gravity,
what is really meant by this is that there are scalar fields (which we collectively denote by $\phi$) with a 
potential $V(\phi)$, such that there is a local (meta)stable de Sitter vacuum 
$\phi=\phi_\star$. The size of the dark energy is then determined dynamically and is given by $V(\phi_\star)$.

In this note I will instead  discuss the quintessence  scenario \cite{Ratra:1987rm,Wetterich:1987fm,Zlatev:1998tr}.
Contrary to the case of the cosmological constant, we are not located at a local minimum of the potential,
and rather the scalar quintessence field $\phi$  is slowly rolling down the potential, even today.

We are interested in quintessence scenarios 
partly because there has been surprisingly relatively few literature on quintessence in quantum-gravity/string-theory contexts 
(see e.g.\ \cite{Choi:1999xn,Choi:1999wv,Hellerman:2001yi,Panda:2010uq,Gupta:2011yj,Cicoli:2012tz} for early literature).
Moreover, in the weak coupling regions of the string-theory landscape
there are some obstructions for constructing de Sitter vacua \cite{Dine:1985he,Maldacena:2000mw,Steinhardt:2008nk,Dvali:2014gua,Dvali:2017eba,Sethi:2017phn,Danielsson:2018ztv,Obied:2018sgi,Andriot:2018wzk,Garg:2018reu,Murayama:2018lie,Ooguri:2018wrx,Hebecker:2018vxz,Garg:2018zdg}. 
While there is no guarantee that we live in such a weakly-coupled region,
this difficulty for de Sitter vacua might be taken as an extra motivation for the quintessence scenario. 
In any case, it ultimately boils down to observation/experiment to tell if any of these scenarios are 
realized in Nature. 

%%%%%%%%%%%%%%%%%%%%%%%%%%%%%%%%%%%%%%%%%%%%%%%%%%%%%%%%%%
\section{Quintessence Axion}
%%%%%%%%%%%%%%%%%%%%%%%%%%%%%%%%%%%%%%%%%%%%%%%%%%%%%%%%%%

One of the questions concerning the quintessence scenario is to explain 
the flatness of the potential---the potential should be very flat, since 
otherwise the size of the dark energy changes too rapidly over time,
and is in contradiction with observational constraints. 
Moreover, such flatness of the potential could easily be spoiled by 
quantum corrections.

One of the possibilities for naturally explaining the flatness of the potential is to impose an
approximate shift symmetry $\phi\to \phi+ (\textrm{const})$.

This approximate symmetry is elegantly realized if the quintessence is 
identified with an ultralight axion (an axion-like particle) \cite{Fukugita:1994hq,Frieman:1995pm,Choi:1999xn}.
Namely, the quintessence field is identified with 
the axion field $a$, which couples to a non-Abelian gauge field $A$ as:
\begin{align}
\mathcal{L}_{aF\tilde{F}}=\frac{a}{32 \pi^2 f}\textrm{Tr} F_{\mu\nu} \tilde{F}^{\mu\nu}\; , 
\label{L_aFF}
\end{align}
where $f$ is the decay constant. Here, we made a crucial assumption that 
 the axion does not have any other tree-level interaction terms beyond the one in Eq.~\eqref{L_aFF}.
 
 The shift symmetry of the axion is broken by non-perturbative effects,
 and the axion potential in the one-instanton approximation is given by
\begin{align}
V(a)=\Lambda^4 \left(1-\cos\frac{a}{f} \right)\; , \label{V_a}
\end{align}
where the size of the dynamical scale $\Lambda$ is naively estimated as
\begin{align}
\Lambda^4=M_{\rm Pl}^4 \, e^{-\frac{2\pi}{\alpha_2}}   \ll M_{\rm Pl}^4 \; .
\label{Lambda_a}
\end{align}
We can explain the size of the dark energy \eqref{Lambda_0} by appropriately 
choosing the value of the gauge coupling constant $\alpha_2$.\footnote{In the quintessence axion scenario,
the energy scale of dark energy is determined dynamically and hence the tracker behavior is not needed, 
contrast to some other quintessence scenarios.}

%%%%%%%%%%%%%%%%%%%%%%%%%%%%%%%%%%%%%%%%%%%%%%%%%%%%%%%%%%
\section{Electroweak Quintessence Axion}
%%%%%%%%%%%%%%%%%%%%%%%%%%%%%%%%%%%%%%%%%%%%%%%%%%%%%%%%%%

While quintessence axion scenario explains the smallness of the 
dark energy, it does not quite explain the particular value for dark energy  \eqref{Lambda_0}:
why should we choose a particular value of $\alpha$? Moreover, it looks that 
this scenario requires an artificial addition of a non-Abelian gauge field to the Standard Model.

It turns out that there is  actually no need to 
add an extra non-Abelian gauge field to the Standard Model,
in order to explain the 
dark energy---the electroweak $\SU(2)$ gauge theory inside the standard model
of particle physics 
does exactly the job. This is the scenario of
electroweak quintessence axion \cite{Nomura:2000yk,McLerran:2012mm}.

One might immediately object that the $\theta$-angle for the electroweak $\SU(2)$ 
gauge group is not physical \cite{Anselm:1992yz}, since $\SU(2)$ theory is chiral and its $\theta$-angle can be rotated away 
by an appropriate $(B+L)$-global symmetry. One expects, however, that the $(B+L)$-global symmetry is broken by the presence of higher dimensional operators, e.g.\ the $qqql/M_{\rm Pl}^2$ operator \cite{Anselm:1993uj}.\footnote{It should also be emphasized that 
there is no exact global symmetry in theories of quantum gravity \cite{Misner:1957mt,Polchinski:2003bq,Banks:2010zn,Harlow:2018tng}.} 

The size of vacuum energy can be estimated by the (constrained) instanton calculus.
Contrary to the case of $\SU(3)$ gauge theory for QCD, in our case the dominant contribution in the 
integral over the size modulus of the instanton comes from small-size instantons---this difference arises 
because as explained above the the instanton contribution is non-zero only when we 
insert $(B+L)$-breaking higher-dimensional operator $qqql/M_{\rm Pl}^2$.
While the small-size instanton contribution naively diverges, 
we expect this to be made finite by some unknown physics near the cutoff scale $M_{\rm Pl}$,
so that the dominant contribution comes from instantons of size $M_{\rm Pl}^{-1}$:\footnote{In this estimate we have taken into account the contribution from the classical instanton action. The one-loop determinant around the instanton solution contributes 
a factor of $(2\pi/\alpha_2(M_{\rm Pl}))^4$, which improves the value to a better value, of order $10^{-120} M_{\rm Pl}^4$.}
\begin{align}
\Lambda^4=M_{\rm Pl}^4 \,  e^{-\frac{2\pi}{\alpha_2(M_{\rm Pl})}} \simeq 10^{-130} M_{\rm Pl}^4  \ll M_{\rm Pl}^4 \;.
\end{align}
Here we use the value of the electroweak coupling constant $\alpha_2=g_2^2/(4 \pi)$ at the Planck scale $\alpha_2(M_{\rm Pl})=1/48$. 
We have neglected numerical factors in this rough estimate.

Note that in evaluating the value of $\alpha_2$ at the cutoff scale we have here implicitly assumed that there are no particles charged under the $\SU(2)$ gauge symmetry
in intermediate energies between electroweak scale and the cutoff scale---we will come back to this assumption in sections \ref{sec.heavy1}
and \ref{sec.heavy2}.

Now, the curious observation is that the
value of $\Lambda$ is close to the current energy scale of the 
dark energy \cite{Nomura:2000yk,McLerran:2012mm} given in \eqref{Lambda_0}!
This numerology is a strong motivation for the electroweak quintessence axion scenario,
where the role of the quintessence axion is played by the axion for the electroweak $\SU(2)$ gauge group.

We still need to make sure that the axion does not roll down the potential too fast. We can estimate this constraint as follows.

If we start with a generic point of the potential, the axion has a mass
$m^2\sim V''\sim \Lambda^4/f^2$. Since the axion starts rolling when the mass is comparable to the Hubble,
we need the condition 
$m>H_0$, where $H_0$ is the present-day value of the Hubble constant:
\begin{align}
m^2&\simeq \frac{\Lambda^4}{f^2} \simeq \frac{H_0^2 M_{\rm Pl}^2}{f^2}
\gtrsim H_0^2 =(2\times 10^{-33} \textrm{eV})^2 \;.
 \label{m}
\end{align}
This constraints can easily be satisfied by a super-Planckian decay constant:\cite{Nomura:2000yk}
\begin{align}
f \gtrsim M_{\rm Pl} \;.
\label{fa_lower}
\end{align}

%%%%%%%%%%%%%%%%%%%%%%%%%%%%%%%%%%%%%%%%%%%%%%%%%%%%%%%%%%
\section{Tension with Weak Gravity Conjecture}
%%%%%%%%%%%%%%%%%%%%%%%%%%%%%%%%%%%%%%%%%%%%%%%%%%%%%%%%%%

The discussion to this point has ween 
within the framework of low-energy effective theory.
However, 
the super-Planckian decay constant \eqref{fa_lower} 
required by the scenario seems to be in tension with known string theory constructions,
    which tends to predict smaller values of the effective decay constant \cite{Banks:2003sx}.
Indeed, 
one of the swampland conjectures, the weak gravity conjecture \cite{ArkaniHamed:2006dz},
sets an upper bound on the decay constant in terms of the gauge coupling constant\footnote{It should be also noted that the value $f\lesssim M_{\rm Pl}$ of the decay constant is also compatible with the refined swampland distance conjecture \cite{Ooguri:2006in,Klaewer:2016kiy}, 
which restricts the field range for the axion to be $\Delta a \lesssim O(M_{\rm Pl})$.}
\begin{align}
f\lesssim \frac{M_{\rm Pl}}{2\pi / \alpha_2(M_{\rm Pl}) } \sim  O(10^{16}\, \textrm{GeV}) \;.
\label{WGC}
\end{align}
The two inequalities, \eqref{fa_lower} and \eqref{WGC}, therefore leaves no allowed value for $f$,
disfavoring the electroweak quintessence axion scenario.

%%%%%%%%%%%%%%%%%%%%%%%%%%%%%%%%%%%%%%%%%%%%%%%%%%%%%%%%%%
\section{Playing Around with Hilltop Quintessence}
%%%%%%%%%%%%%%%%%%%%%%%%%%%%%%%%%%%%%%%%%%%%%%%%%%%%%%%%%%

Is there any way around this problem?

If one takes into the constraints from the weak gravity conjecture \eqref{WGC} seriously, 
one can try to change the estimate \eqref{fa_lower}, so that we can lower the value of the decay constant.
This can be achieved by fine-tuning the initial condition to 
a local maximum $a\sim \pi f$ of the potential.
In this case, the quintessence has already started rolling down the potential,
however is still located close to the local maximum. 

This hilltop quintessence scenario \cite{Dutta:2008qn}, however, has a severe
fine-tuning problem in our context. Namely, we choose do the 
$\exp(O(M_{\rm Pl}/ f))=\exp(O(100))$ fine-tuning for the 
 initial value $\delta a=a-\pi f$ \cite{Choi:1999xn,Ibe:2018ffn}!
 One moreover expects that such an extreme fine-tuning is incompatible with quantum 
fluctuations generated during the inflationary era. This is a strong argument against hilltop fine-tuning.

%%%%%%%%%%%%%%%%%%%%%%%%%%%%%%%%%%%%%%%%%%%%%%%%%%%%%%%%%%
\section{Playing Around with Heavy Particles}\label{sec.heavy1}
%%%%%%%%%%%%%%%%%%%%%%%%%%%%%%%%%%%%%%%%%%%%%%%%%%%%%%%%%%

When one looks back at the tension between the two inequalities \eqref{fa_lower} and \eqref{WGC},
one notices that the tension arises due to a small value of 
$\alpha_2(M_{\rm Pl})$. We computed this by assuming that there is 
no $\SU(2)$-charged particles between the electroweak scale and the Planck scale.
However, there is apriori no strong justification for assumption.

This motivates us the following scenario:
what happens if there are heavy particles in the intermediate energy scales,
such that the RG running of the  gauge coupling constant $\alpha_2$
changes sufficiently, so that its value at the Planck scale is large: $\alpha_2(M_{\rm Pl})\sim 2\pi$?
If this happens, then the constraints from the weak gravity conjecture \eqref{WGC} now reads
$f\lesssim M_{\rm Pl}$, so that the two conditions \eqref{fa_lower} and \eqref{WGC}
are satisfied by the value $f \sim M_{\rm Pl}$.

Of course, by choosing  $\alpha_2/(2\pi) \sim O(1)$
we are in a strongly-coupled regime, where the 
potential in the one-instanton approximation \eqref{V_a} can
no longer be trusted. However, this is not a problem for us, since 
nowhere in our argument we used the particular cosine form of the potential.
What is important for explanation of dark energy is that 
we have a non-trivial potential, whose size is given by the dynamical scale.
This is still expected to be the case in the strongly-coupled regime.

Unfortunately, this idea of changing the RG flow turns out be 
fundamentally flawed. The reason is that since once we change $\alpha_2(M_{\rm Pl})$
we also change the dynamical scale \eqref{Lambda_a}, 
spoiling the successful numerological coincidence $\Lambda \simeq \Lambda_0$, 
which was the very starting point for our quintessence axion scenario.

%%%%%%%%%%%%%%%%%%%%%%%%%%%%%%%%%%%%%%%%%%%%%%%%%%%%%%%%%%
\section{Supersymmetric Extension}
%%%%%%%%%%%%%%%%%%%%%%%%%%%%%%%%%%%%%%%%%%%%%%%%%%%%%%%%%%

Interestingly, the problems discussed above are solved elegantly in the minimal supersymmetric extension of the standard model (MSSM),
which we turn next. We assume that the supersymmetry is spontaneously broken, and we denote
the sparticle mass scale by $m_{\rm SUSY}$. 

In the MSSM, the $B+L$ symmetry is broken by Planck-suppressed dimension $5$ operators $QQQL$  \cite{Sakai:1981pk,Weinberg:1981wj}.
These operators induce too rapid proton decay for $m_{\rm SUSY}\simeq O(\textrm{TeV})$.
To suppress the dimension $5$ operators, we assume Froggatt-Nielsen (FN) symmetry
\cite{Froggatt:1978nt}. 
We also assume the global $R$-symmetry given in  \cite{Nomura:2000yk}.
The FN symmetry is spontaneously broken by a VEV of a scalar field $\phi$ (with a small VEV $\langle \phi \rangle / M_{\rm Pl}=\epsilon\ll 1$), where the value of $\epsilon$ is taken to be 
$\epsilon \simeq 1/17$ \cite{Buchmuller:1998zf} to explain the quark/lepton masses and mixing angles.
With the charge assignment in  Table 2 of \cite{Nomura:2000yk}, we confirm that the 
proton decay via the $QQQL$ operators are sufficiently suppressed.

In addition to suppressing dangerous dimension $5$ operators and 
explaining the quark/lepton flavor structures,
FN symmetry is also of help in
getting a magnitude for the dark energy. One might naively expect that the 
estimate for the dynamical scale $\Lambda$ in MSSM is 
changed from \eqref{Lambda_a} into 
\begin{align}
\Lambda^4 
\overset{?}{\simeq} 
e^{-\frac{2\pi}{\alpha_2^{\rm MSSM}(M_{\rm Pl}) }} M^4_{\rm Pl} \; , \label{Lambda_a_wrong}
\end{align}
where $\alpha_2^{\rm MSSM}(M_{\rm Pl})$ is the $\SU(2)$ gauge coupling constant at the Planck scale for the MSSM---if do we not include any heavy particles beyond the MSSM, we can substitute the value $\alpha_2^{\rm MSSM}(M_{\rm Pl})\simeq 1/23$,
which is too large for dark energy.

One notices, however, that the estimate \eqref{Lambda_a_wrong} is flawed. Indeed, we need supersymmetry breaking for a non-zero value of $\Lambda$, and hence the correct answer should be suppressed by a suitable power of $m_{\rm SUSY}$.
Also, since we need higher-dimensional $(B+L)$-breaking operators the answer should also we suppressed by 
a suitable power of $\epsilon$.

The correct answer, taking into account these points, is given by\cite{Nomura:2000yk}\footnote{The powers of $\epsilon$ and $m_{\rm SUSY}$ can be computed by an explicit instanton computation. Alternatively, the powers can be computed via the mixed anomaly between
$SU(2)$ electroweak gauge theory and $U(1)$ FN or $U(1)$ R-symmetry.}
\begin{align}
\Lambda^4
\simeq e^{-\frac{2\pi}{\alpha^{\rm MSSM}_2(M_{\rm Pl}) }}  \epsilon^{10} m_{\rm SUSY}^3 M_{\rm Pl}  
\simeq \left(\frac{\epsilon}{1/17}\right)^{10} \left(\frac{m_{\rm SUSY}}{1 \textrm{ TeV}} \right)^3
(O(\textrm{meV}))^4 \; .  \label{Lambda_a_SUSY}
\end{align}
Choosing the value $\epsilon \simeq 1/17$ for quark/lepton mixing matrices and 
the supersymmetry breaking scale to be $m_{\rm SUSY} \simeq 1 \textrm{ TeV}$,
one indeed finds the correct size for dark energy.

%%%%%%%%%%%%%%%%%%%%%%%%%%%%%%%%%%%%%%%%%%%%%%%%%%%%%%%%%%
\section{Supersymmetric Miracle}\label{sec.heavy2}
%%%%%%%%%%%%%%%%%%%%%%%%%%%%%%%%%%%%%%%%%%%%%%%%%%%%%%%%%%

Suppose that  we include a pair of heavy massive particle $X, \bar{X}$ 
in some representation $R$ of $\SU(2)$ gauge group.
When the mass of $X, \bar{X}$ is at an intermediate energy scale $M_X$,
then the RG running of the coupling constant in the one-loop approximation is modified as
\begin{align}
\alpha_2^{-1} (M_{\rm Pl})|_{X\bar{X}}=\alpha_2^{-1} (M_{\rm Pl})\big|_{\rm MSSM} + \frac{2 T_R}{2 \pi} \log \frac{M_X}{M_{\rm Pl}}\; , 
\end{align}
where $T_R$ is the Dynkin index of the representation $R$.

Such heavy particles, however also change the 
zero modes---a chiral multiplet always contains a fermion.
This means we need to insert operators $M_X X \bar{X}$ to
cancel the instanton zero modes, leading to an extra factor
$(M_X/M_{\rm Pl})^{2T_R}$.
 
Interestingly, these two effects cancel with each other for a supersymmetric theory,
leaving the dynamical scale $\Lambda$ invariant \cite{Nomura:2000yk}:
\begin{align}
\Lambda^4|_{X\bar{X}}
&\simeq e^{-\frac{2\pi}{\alpha_2(M_{\rm Pl})|_{X\bar{X}} }} 
\left(\frac{M_X}{M_{\rm Pl}}\right)^{2T_R} \epsilon^{10} m_{\rm SUSY}^3 M_{\rm Pl} 
= c\, e^{-\frac{2\pi}{\alpha_2(M_{\rm Pl})|_{\rm MSSM } }}  \epsilon^{10} m_{\rm SUSY}^3 M_{\rm Pl} \simeq \Lambda\big|_{\rm MSSM}^4 \; .\label{Lambda_preserved}
\end{align}
Due to the miraculous cancellation in supersymmetric theory,
$\alpha_2(M_{\rm Pl})$ 
can be achieved while keeping the relation $\Lambda\simeq \Lambda_0$ intact.\footnote{While the one-instanton approximation is not reliable for a large value of the coupling constant $\alpha_2(M_{\rm Pl})\sim 2\pi$,
we expect that the estimation of the energy scale $\Lambda$ will not be 
significantly affected by this subtlety.} This solves the tension mentioned above.
We can say, to the least, that the tension between flatness of the potential \eqref{fa_lower} and 
the weak gravity conjecture \eqref{WGC} is significantly ameliorated in the 
supersymmetric setup.

One should also notice that the energy scales of the dark energy
is now much more robust in the supersymmetric case.
As we discussed in section \ref{sec.heavy1}, 
the successful prediction for the energy scale of dark energy can 
be spoiled if there are heavy particle in intermediate energy scales.
Our scenario ensures that the energy scale is kept intact in 
the supersymmetric case, as long as supersymmetry is already restored in the
intermediate energy scale in question and if the heavy particles in question 
are in the supersymmetric multiplet. 

There are many scenarios for realizing $\alpha_2(M_{\rm Pl})\sim 2\pi$ (so that $f\simeq M_{\rm Pl}$).
For example, we can simply include $3$ $\SU(2)$ triplets at $\sim 10^7\, \textrm{GeV}$.
As another possibility consistent with a coupling unification as in grand unified theories (GUT), 
let us consider an $\SU(2)$ triplet and an $\SU(3)$ octet at $10^{12} \textrm{GeV}$.
We see that all gauge couplings meet at the Planck scale \cite{Bachas:1995yt,Bhattacharyya:2013xba}.
The value $\alpha_2(M_{\rm Pl})\sim 2\pi$ can then be achieved by 
including $4$ pairs of GUT-like multiplets $\bf{5}$ and $\bar{\bf{5}}$ at $\sim 10\, \textrm{TeV}$.
In any of those scenarios, $\Lambda \sim \Lambda_0$ is not affected by the Supersymmetric Miracle.

\section{Summary}

In this short note we discussed the electroweak quintessence axion as a candidate for dark energy.
In this scenario, the correct size for 
dark energy can naturally be explained by the 
dynamics of the electroweak $\SU(2)$ gauge theory.
It is interesting that the $\theta$-angle for the $\SU(2)$ gauge group,
which is often not included into the parameters for the Standard Model,
provides a minimalistic scenario for dark energy.

Our discussion combines inputs from observations, Standard Model of particle physics,
as well as swampland constraints. These considerations have lead us to 
the supersymmetric extension of the scenario,
where constraints from weak gravity conjecture can be satisfied
without fine-tuning into the hilltop region. We also pointed out that
the energy scale of dark energy is robust again the possible presence of 
supersymmetric heavy matters in intermediate energy scales.

Let us emphasize here that we have not claimed to have solved the cosmological constant problem in full generality.
The small energy scale generated by the electroweak axion can in general be overshadowed easily by 
zero-point energies of various fields present in the standard model and elsewhere. 
Nevertheless, understanding the dynamical origin of the correct order-of-estimate for the 
dark energy is likely be a crucial step in this direction.

The electroweak quintessence axion is a compelling idea.
On the one hand, 
it is a fascinating question to explore if our scenario can really be realized in a 
specific setup inside theories of quantum gravity, such as string theory (see e.g.\ \cite{Hebecker:2019csg
} for recent discussion).
On the other hand, this scenario should be tested by future observations,
such as LSST, Euclid and WFIRST. Overall, it is clear that we are presently at an exciting time for dark energy, both 
theoretically and observationally.

\section*{Acknowledgments}

The author would like thank ICISE (rencontres du Vietnam, 2019) and PASCOS 2019 (Manchster, 2019) for providing a stimulating environment. The author is partially supported by WPI program (MEXT, Japan) and by JSPS KAKENHI Grant No. 17KK0087, No. 19K03820 and No. 19H00689.

%%%%%%%%%%  Bibliography  %%%%%%%%%%%%
\bibliographystyle{nb}  
\bibliography{swampland_Vietnam}
%%%%%%%%%%%%%%%%%%%%%%%%%%%
%%%%%%%%%%%%%%%%%%%%%%%%%%%%
\end{document}